\documentclass[aps,prb,twocolumn,preprintnumbers,amsmath,amssymb,superscriptaddress]{revtex4}%

\usepackage{graphicx}%
\usepackage{dcolumn}
\usepackage{amsmath}
\usepackage{color}
\usepackage{ulem}

\begin{document}

\title{Magnetic states of MnP: muon-spin rotation studies}
\author{R.~Khasanov}
 \email{rustem.khasanov@psi.ch}
 \affiliation{Laboratory for Muon Spin Spectroscopy, Paul Scherrer Institut, 5232 Villigen PSI, Switzerland}
\author{A.~Amato}
 \affiliation{Laboratory for Muon Spin Spectroscopy, Paul Scherrer Institut, 5232 Villigen PSI, Switzerland}
\author{P.~Bonf\`{a}}
 \affiliation{Dipartimento di Fisica e Scienze della Terra e  Unit\`{a} CNISM di Parma, Universit\`{a} di Parma, 43124 Parma, Italy}
\author{Z.~Guguchia}
 \affiliation{Laboratory for Muon Spin Spectroscopy, Paul Scherrer Institut, 5232 Villigen PSI, Switzerland}
\author{H.~Luetkens }
 \affiliation{Laboratory for Muon Spin Spectroscopy, Paul Scherrer Institut, 5232 Villigen PSI, Switzerland}
\author{E.~Morenzoni}
 \affiliation{Laboratory for Muon Spin Spectroscopy, Paul Scherrer Institut, 5232 Villigen PSI, Switzerland}
\author{R.~De~Renzi}
 \affiliation{Dipartimento di Fisica e Scienze della Terra e  Unit\`{a} CNISM di Parma, Universit\`{a} di Parma, 43124 Parma, Italy}
\author{N.D.~Zhigadlo}
 \affiliation{Laboratory for Solid State Physics, ETH Zurich, 8093 Zurich, Switzerland,}
 \affiliation{Department of Chemistry and Biochemistry, University of Bern, 3012 Bern, Switzerland}

\begin{abstract}
Muon-spin rotation data collected at ambient pressure ($p$) and at $p=2.42$~GPa in MnP were analyzed to check their consistency with various low- and high-pressure magnetic structures reported in the literature.   Our analysis confirms that in MnP the low-temperature and low-pressure helimagnetic phase is characterised by an increased value of the average magnetic moment compared to the high-temperature ferromagnetic phase. An elliptical double-helical structure with a propagation vector ${\bf Q}=(0,0,0.117)$, an $a-$axis moment elongated by approximately 18\% and an additional tilt of the rotation plane towards $c-$direction by $\simeq 4-8^{\rm o}$ leads to a good agreement between the theory and the experiment.
The analysis of the high-pressure $\mu$SR data reveals that the new magnetic order appearing for pressures exceeding $1.5$~GPa can not be described by keeping the propagation vector ${\bf Q} \parallel c$. Even the extreme case -- decoupling the double-helical structure into four individual helices -- remains inconsistent with the experiment. It is shown that the high-pressure magnetic phase which is a precursor of superconductivity is an incommensurate helical state with ${\bf Q} \parallel b$.
\end{abstract}
\maketitle

\section{Introduction}

Recently, the helical magnets CrAs and MnP  have attracted much interest due to the discovery of superconductivity induced by hydrostatic pressure.\cite{Wu_NatConn_2014, Kotegawa_JPSJ_2014, Kotegawa_PRL_2015, Cheng_PRL_2015} CrAs is fully magnetic below the N\'{e}el temperature $T_{\rm N}$ up to pressures of $p\simeq 0.4$~GPa, phase separated between  magnetic and nonmagnetic volumes (with the latter ones becoming superconducting below $\simeq 2$~K) for $0.4\lesssim p\lesssim 0.7$~GPa and purely superconducting below the transition temperature $T_c$ for pressures exceeding 0.7~GPa.\cite{Khasanov_SciRep_2015} It is important to emphasize here that in CrAs a single type of double-helical magnetic order remains {\it unchanged} as a function pressure.\cite{Khasanov_SciRep_2015}

The binary pnictide MnP possesses a more complicated phase diagram.\cite{Cheng_PRL_2015} At ambient pressure MnP orders ferromagnetically at $T\simeq 290$~K  with the Mn magnetic moments aligned along the crystallographic $b-$direction (according to the crystallographic group $Pnma$ 62 with lattice constants $c>a>b$).\cite{Huber_PR_1964, Felcher_JAP_1966, Yamazaki_JPSJ_2014}  The ordered moment is $m\simeq 1.29$~$\mu_{\rm B}$ per Mn atom.\cite{Huber_PR_1964, Obara_JPSJ_1980} At lower temperatures ($T\lesssim 50$~K) another transition into a double-spiral helical structure is reported.\cite{Felcher_JAP_1966, Obara_JPSJ_1980, Forsyth_PPS_1966} In this helimagnetic state (helical$-c$ state, Ref.~\onlinecite{Matsuda_Arxiv_2016}) the Mn moments rotate within the $ab$-plane (helical plane) with the propagation vector ${\bf Q}=(0,0,0.117)$ normal to the helical plane.\cite{Felcher_JAP_1966} Following Forsyth {\it et al.}\cite{Forsyth_PPS_1966}, the average magnetic moment in the helical state increases up to $m\simeq 1.58$~$\mu_{\rm B}$ with a longer component ($m_b=1.73$~$\mu_{\rm B}$) along the $b-$axis and a shorter one ($m_a=1.41$~$\mu_{\rm B}$) along the $a-$  axis, respectively. This contradicts conclusions of Obara {\it et al.}\cite{Obara_JPSJ_1980} who obtain $m\simeq 1.33$~$\mu_{\rm B}$ in both, the ferromagnetic and the helical$-c$ states, and claim a  'circular' helical$-c$ state with $m_a=m_b$. In addition, the recent work of Yamazaki {\it et al.}\cite{Yamazaki_JPSJ_2014} reports the observation of a new magnetic reflection at $(1,0,\delta)$,  indicating that the helical planes are tilted towards the $c-$direction alternately.

A hydrostatic pressure of about 2~GPa induces a transition from the double-helical structure to a different type of antiferromagnetic ground state below $T_{\rm N}\simeq 150$~K. The nature of the new magnetic order, which precedes the appearance of superconductivity, is still puzzling. Two possible scenarios were considered based on results of x-ray, neutron scattering and muon-spin-rotation ($\mu$SR) experiments,\cite{Wang_Arxiv_2015, Matsuda_Arxiv_2016, Khasanov_PRB_2016} as well as the ones predicted theoretically based on first principal calculations.\cite{Bonfa_Arxiv_2016} The first, advocated in Ref.~\onlinecite{Wang_Arxiv_2015}, identifies the new state still as a helical one, albeit with an increased propagation vector of ${\bf Q} \simeq (0, 0, 0.25)$ compare to the ambient pressure value of ${\bf Q}=(0,0,0.117)$. The second, Refs.~\onlinecite{Matsuda_Arxiv_2016, Khasanov_PRB_2016, Bonfa_Arxiv_2016}, discusses the change of the propagation vector from ${\bf Q} \parallel c$ to ${\bf Q} \parallel b$.

The main purpose of this paper is to present a consistency check of the recently published muon-spin rotation data, Ref.~\onlinecite{Khasanov_PRB_2016}, with the various low- and high-pressure magnetic phases of MnP that were reported in Refs.~\onlinecite{Felcher_JAP_1966, Obara_JPSJ_1980, Forsyth_PPS_1966, Yamazaki_JPSJ_2014, Wang_Arxiv_2015, Matsuda_Arxiv_2016, Khasanov_PRB_2016}. Our analysis appears to confirm that in MnP the low-pressure helimagnetic phase is indeed characterised by an increased  value of the average magnetic moment in comparison with the ferromagnetic one. The elliptical helical structure with ${\bf Q}=(0,0,0.117)$, an $a-$axis moment elongated by approximately 18\%, as well as an additional tilt of the rotation plane towards $c-$direction by  $\simeq 4-8$ degree leads to a reasonably good agreement between the theory and the experiment.
The analysis of the high-pressure $\mu$SR data reveals that the new magnetic order appearing for pressures exceeding $1.5$~GPa can not be described by keeping the direction of the propagation vector unchanged. An agreement between the contradicting neutron and $\mu$SR data on the one side and the x-ray data on the other side may be reached by assuming a tilt of the rotation plane along the $c-$direction which leads to the appearance of a $b-$axis modulation of the Mn magnetic moments.

The paper is organized as follows. In Sec.~\ref{sec:experimental} the sample preparation details, the pressure cell description and a short overview of the performed $\mu$SR experiments are given. Section~\ref{sec:muon-stopping} describes the MnP unit cell and gives a brief description of the calculation of the muon stopping sites and the local fields seen by the muons. A comparison of ambient pressure $\mu$SR data with ferromagnetic and helical$-c$ types of the magnetic orders is given in Sec.~\ref{sec:ambient-pressure}. Sec.~\ref{sec:high-pressure} comprises studies of the high-pressure magnetic state of MnP. Conclusions follow in Sec.~\ref{sec:conclusions}.

\section{Experimental} \label{sec:experimental}

\subsection{Sample preparation}

The manganese phosphide (MnP) polycrystalline sample was synthesized by using a high-pressure furnace. Overall details of the sample cell assembly and high-pressure synthesis process can be found in Ref.~\onlinecite{Zhigadlo_PRB_2012}. The mixture of manganese powder (99.99\%) and red phosphorus powder (99.999\%) in a molar ratio 1:1 was enclosed in a boron-nitride crucible and placed inside a pyrophyllite cube with a graphite heater. All the preparatory steps were done in a glove box under argon atmosphere. In a typical run, the sample was compressed up to 1~GPa at room temperature. While keeping the pressure constant, the temperature was ramped up within 3~h to the maximum value of 1200~$^{\rm o}$C, kept stable for 1~h, decreased to 950~$^{\rm o}$C within 14~h and finally quenched to the room temperature. Afterwards, the pressure was released and the final solid product removed. Subsequently recorded x-ray powder diffraction patterns showed no secondary phases.

\subsection{Experimental techniques}

\subsubsection{Pressure Cell}

The pressure was generated in a double-wall piston-cylinder type of cell made of MP35N alloy. As a pressure transmitting medium 7373 Daphne oil was used. The pressure was measured in situ by monitoring the pressure shift of the superconducting transition temperature of In. The details of the experimental setup for conducting $\mu$SR under pressure experiments are provided in Ref.~\onlinecite{Khasanov_HPR_2016}.

\subsubsection{Muon-spin rotation}

$\mu$SR measurements were  performed at the $\pi$M3 and $\mu$E1 beamlines (Paul Scherrer Institute, Villigen, Switzerland), by using the GPS and GPD spectrometers,\cite{Khasanov_HPR_2016} respectively. At the GPS spectrometer, equipped with a continuous flow $^4$He cryostat, zero-field (ZF) $\mu$SR experiments at ambient pressure and down to temperatures of $1.6$~K  were carried out. At the GPD spectrometer, equipped with a continuous flow $^4$He cryostat (base temperature $\simeq 2.2$~K), ZF-$\mu$SR experiments under pressure up to $\sim$2.4~GPa were conducted.

\section{Internal fields at the muon stopping sites} \label{sec:muon-stopping}

\subsection{MnP unit cell}

\begin{figure}[htb]
\includegraphics[width=0.8\linewidth]{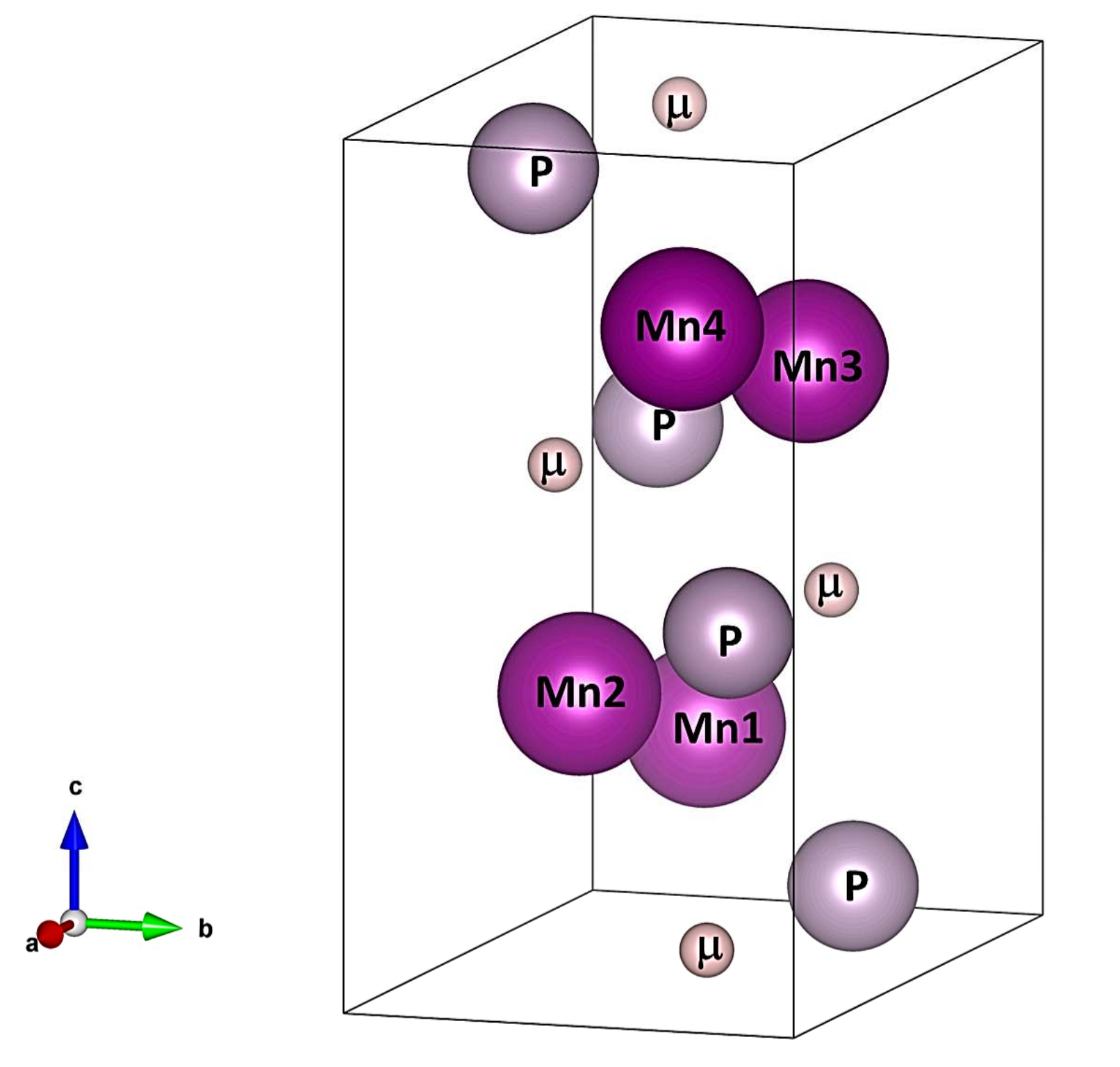}
%
\caption{The orthorhombic crystal structure of MnP ($Pnma$, 62). The muon stopping positions (sites \#1)  were obtained by $ab-initio$ calculations (see Sec.~\ref{sec:stopping-sites} and Table~\ref{tab:dftmuontable}). The structure  was visualized by using VESTA.\cite{Vesta}
}
 \label{fig:MnP_Unit-cell}
\end{figure}

The orthorhombic crystal structure of MnP ($Pnma$, 62) is shown in Fig.~\ref{fig:MnP_Unit-cell}. The unit cell dimensions are $a=5.268$~\AA, $b=3.172$~\AA\ and $c=5.918$~\AA\  at ambient pressure and room temperature. Both the Mn and the P atoms occupy the $4c$ $(x, 1/4, z)$ crystallographic positions with $x_{\rm Mn} = 0.0049(2)$, $z_{\rm Mn} = 0.1965(2)$ and $x_{\rm P} = 0.1878(5)$, $z_{\rm P} = 0.5686(5)$.\cite{Rundqvist_ACS_1962}

\subsection{Muon stopping sites}\label{sec:stopping-sites}

The {\it ab initio} identification of the muon stopping sites was performed with the method described in Ref.~\onlinecite{jp5125876}. The description of the electronic density was obtained with DFT using a plane wave and pseudopotential approach as implemented in the {\sc Quantum ESPRESSO} suite of codes.\cite{QE-2009} The reciprocal space was sampled with a  $6 \times 8 \times 12$ Monkhorst-Pack  grid.\cite{PhysRevB.13.5188} The exchange-correlation functional of Perdew, Burke, and Ernzerhof and the Methfessel-Paxton scheme with 0.01 Ry smearing were  used.\cite{PhysRevLett.77.3865,PhysRevLett.78.1396,PhysRevB.40.3616} The ultrasoft pseudopotentials described in Ref.~\onlinecite{Garrity2014446} and a basis set expanded up to a kinetic-energy cutoff of 70 Ry and up to 500 Ry for
charge density were adopted.
These settings guarantee an accurate description of the crystalline structure  of the material. In the collinear spin formalism, at ambient pressure, the ferromagnetic state has the lowest enthalpy and it is therefore considered as the ground state for the structural relaxations of the impurity in the suprecells.

A supercell containing 129 atoms (including the muon which is described as a hydrogen atom) is used to locate the  possible interstitial embedding positions occupied by the muon. To get a reasonable compromise between speed and accuracy, the kinetic energy cutoff and the charge density cutoff were reduced to 60 Ry and 400 Ry respectively. The Baldereschi point $\mathbf{k}=({1}/{4},{1}/{4},{1}/{4})$  was used to sample the reciprocal space.\cite{PhysRevB.7.5212}
The lattice cell parameters were kept  fixed during the relaxation.

A grid of $4 \times 4 \times 4$ initial interstitial positions  was selected to explore the whole interstitial space of the unit cell and identify all the possible embedding sites. After removing the positions too close to the atoms of the hosting system (less than 1~{\AA}) and disregarding symmetry equivalent sites, a set of 9 interstitial locations was obtained. The structural relaxations were performed with the convergence criteria set to $10^{-4}$ Ry for the total energy and to $10^{-3}$ Ry/a.u. for forces. Five interstitial positions have been identified with this procedure.
The total energy differences between the possible interstitial sites is reported in Table \ref{tab:dftmuontable}.

\begin{table}
\begin{tabular}{cccc}
\hline \hline
muon sites & Position & muons per u.c.&$\Delta E $\\
&&&(eV)\\
\hline
\#1 & (0.103, 0.25, 0.921)  & 4& 0  \\
\#2 & (0.486, 0.935, 0.938) & 8&0.8  \\
\#3 & (0.549, 0.75, 0.954)  & 4&0.8  \\
\#4 & (0.93, 0.25, 0.726)   & 4&0.3  \\
\#5 & (0.772, 0.04, 0.672)  & 8& 0.4   \\
\hline \hline
\end{tabular}
\caption{List of the candidate muon embedding sites identified with
supercell structural relaxations. The energy differences $\Delta E=E_i-E_{\#1}$ are referred to sites \#1. The ''muons per u.c.`` referes to the number of equivalent muon positions within the unit cell. }\label{tab:dftmuontable}
\end{table}

The identification of multiple candidate sites is not an unexpected feature of DFT based muon site
assignments.\cite{jp5125876,PhysRevLett.114.017602,PhysRevB.91.144417,Moller_PRB_2013,PhysRevB.88.064423} This is partially caused by the structural optimization algorithm which neglects both the zero point motion energy of the muon and the effects of temperature. Molecular dynamics approaches would substantially improve the accuracy of the results, but they would also result in a tremendous increase of the computational costs. The selected convergence criteria may also cause the relaxation algorithm to stop in configurations which are not real minima but rather constitute a flat area between different interstitial positions.

Since only one frequency is observed in the FM phase of MnP, it is reasonable to assume that only one fully populated type of muon sites is present in this material. Simulations performed with the double adiabatic approximation and the
exploration algorithm discussed in Ref.~\onlinecite{jp5125876} show that the energy barrier binding the muon in  sites \#1 is larger than 0.5 eV while the same analysis shows that  sites \#4 and \#5 cannot bind a muon since their energy barriers are of the order of 0.1 eV. Sites \#1 were therefore selected for the subsequent analysis of the experimental data (see Fig.~\ref{fig:MnP_Unit-cell}).

\subsection{Local field at the muon stopping site}

Muons probe the local field, which is the vector sum of the internal (dipolar) magnetic field and the contact field at a particular site. The spontaneous local field for the site $i$ was calculated as:
\begin{equation}
{\bf B}_{{\rm loc},i}={\bf B}_{{\rm dip},i}+{\bf B}_{{\rm cont},i}
 \label{eq:Bloc}
\end{equation}

The dipolar magnetic field $B_{\rm dip}({\bf r})$ at position ${\bf r}$ within the lattice unit cell was:\cite{Blundell_PhysB_2009}
\begin{equation}
B_{\rm dip}^\alpha({\bf r})=\frac{\mu_0}{4\pi} \sum_{j,\beta} \frac{m_j^\beta}{R_j^3}
 \left( \frac{3 R_j^\alpha R_j^\beta}{R_j^2} - \delta^{\alpha \beta} \right)
 \label{eq:Bdip}
\end{equation}
Here ${\bf R}_j={\bf r}-{\bf r}_j$, $\alpha$ and $\beta$ denote the vector components $x$, $y$, and $z$, ${\bf r}_j$ is the position of $j-$th magnetic ion in the unit cell, and $m_j^\beta$ is the corresponding dipolar moment. The summation is taken over a sufficiently large Lorentz sphere of radius $R_L$.

The contact field ${\bf B}_{\rm cont}$ was obtained as:
\begin{equation}
{\bf B}_{{\rm cont},i}=A_{\rm cont}\sum_{j=1}^N \omega(j){\bf m}_j,
 \label{eq:Bcont}
\end{equation}
where $A_{\rm cont}$ is the contact coupling constant, ${\bf m}_j$ are $N$ nearest neighboring magnetic moments and $\omega(j)$ is the weight obtained as $\omega(j)=R_j^{-3}/\sum_{k=1}^N R_k^{-3}$ with $R_k$ being the distance between the $i$-th muon and $k$-th magnetic moment.

The contact field $ B_{\rm cont}$ was calculated by using Eq.~\ref{eq:Bcont} and considering 3 nearest neighbours Mn ions for each particular muon site.

\section{Ambient pressure magnetism}\label{sec:ambient-pressure}

\subsection{Ferromagnetic state}

\begin{figure}[htb]
\includegraphics[width=1.0\linewidth]{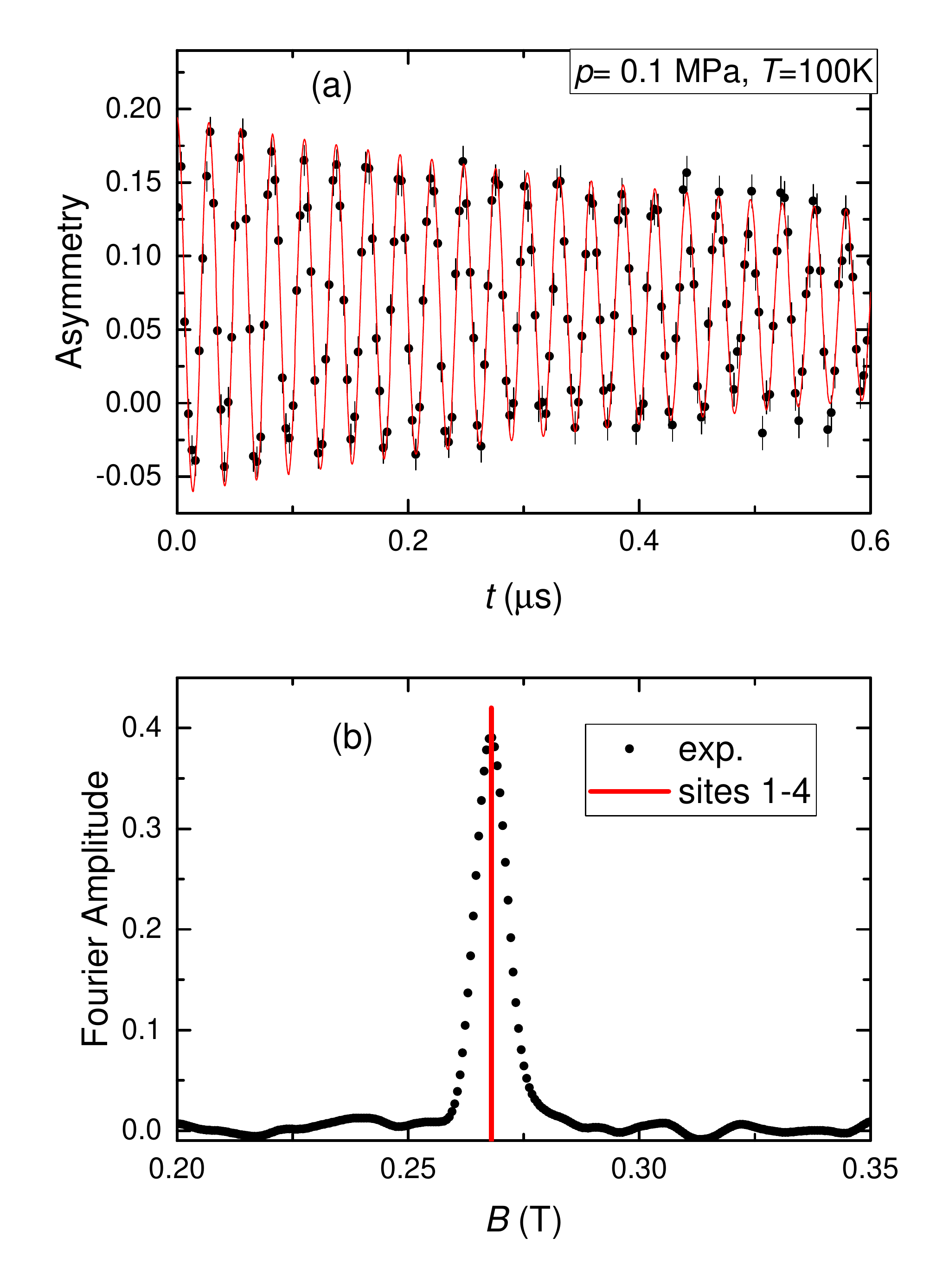}
%
\caption{(a) The ZF-$\mu$SR time spectrum of MnP at ambient pressure in the ferromagnetic state ($p=0.1$~MPa, $T=100$~K). The solid line is the fit using Eq.~\ref{eq:P_FM}. (b) The fast Fourier transform of the ZF-$\mu$SR time spectrum. The solid line corresponds to the local field $B_{\rm loc}$ on the four muon sites \#1 (see text for details).
}
 \label{fig:FM}
\end{figure}

A zero-field muon time spectrum of MnP recorded at ambient pressure in the ferromagnetic (FM) state ($p=0.1$~MPa, $T=100$~K) is presented in Fig.~\ref{fig:FM}a. The corresponding fast Fourier transform of the $\mu$SR time spectrum is shown in Fig.~\ref{fig:FM}b.

The solid line in Fig.~\ref{fig:FM}a corresponds to a fit with the following function:
\begin{equation}
\frac{A^{\rm FM}(t)}{A^{\rm FM}(0)}=P^{\rm FM}(t)=\frac{2}{3}e^{-\lambda_T t} \cos(\gamma_\mu
B_{int}t) + \frac{1}{3}\;e^{-\lambda_L t}.
 \label{eq:P_FM}
\end{equation}
Here $P^{\rm FM}$ is the time dependent muon-spin polarization, $A^{\rm FM}(0)$ is the initial asymmetry,  $B_{int}$ is the internal field at the muon stopping site, $\gamma_{\mu}= 2\pi\;135.5$~MHz/T is the muon gyromagnetic ratio, and $\lambda_T$ and $\lambda_L$ are the transverse and the longitudinal exponential relaxation rates, respectively. The occurrence of 2/3 oscillating and 1/3 non-oscillating $\mu$SR signal fractions originates from the spatial averaging in powder samples, where 2/3 of the magnetic field components are perpendicular to the muon-spin and cause a precession, while the 1/3 longitudinal field components do not.\cite{Yaouanc_book_11}

The calculations of the local fields at muon sites were first performed by assuming the known ferromagnetic structure and an ordered moment of $m\simeq 1.29$~$\mu_{\rm B}$ per Mn atom.\cite{Huber_PR_1964, Felcher_JAP_1966, Obara_JPSJ_1980, Forsyth_PPS_1966} As a first step, the dipolar fields ($B_{\rm dip,i}$) for the four crystallographically equivalent muon sites \#1 (see Fig.~\ref{fig:MnP_Unit-cell} and Tab.~\ref{tab:dftmuontable}) were calculated by means of Eq.~\ref{eq:Bdip}. All four fields were found to be the same (as one can expect for a FM structure). As a second step, from the calculated $B_{\rm dip}$ and the measured $B_{\rm int}$ the contact field $B_{\rm cont}$ is obtained from Eq.~\ref{eq:Bloc}. Finally, from the known value of $B_{\rm cont}$ and by using Eq.~\ref{eq:Bcont}, the coupling contact constant $A_{\rm cont}\simeq -0.474$~${\rm T}/\mu_{\rm B}$ was determined.\cite{comment}

\subsection{Helical$-c$ state} \label{sec:HelC}

\begin{figure}[htb]
\includegraphics[width=1.0\linewidth]{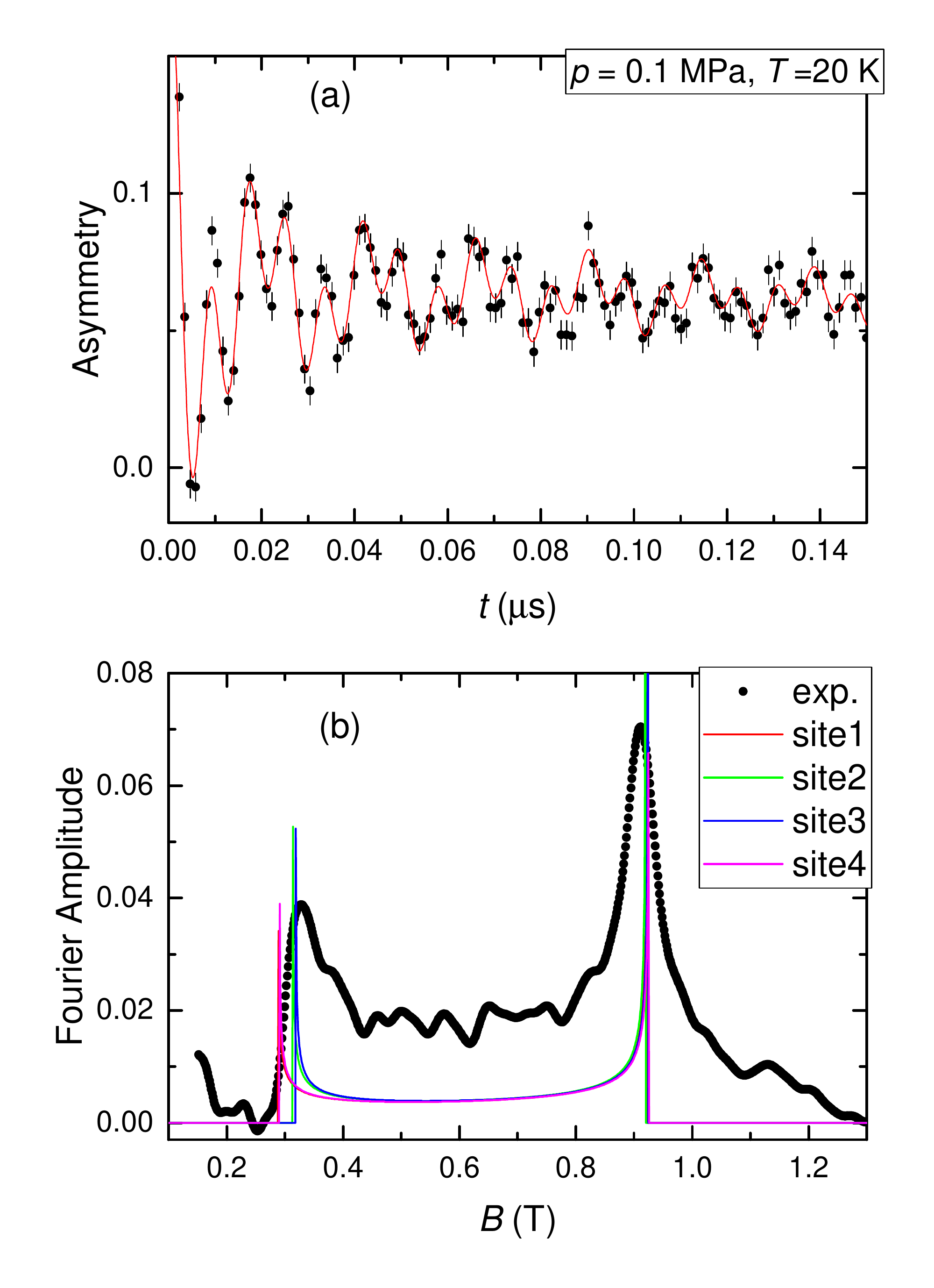}
%
\caption{(a) The ZF-$\mu$SR time spectrum of MnP at ambient pressure in the helical$-c$ state ($p=0.1$~MPa, $T=20$~K). The solid line corresponds to the fit of Eq.~\ref{eq:P_Hel} to the data with $\beta=6^{\rm o}$, $m_a=1.27$~$\mu_{\rm B}$,  $m_b=1.83$~$\mu_{\rm B}$ and $\alpha=16^{\rm o}$. (b) The fast Fourier transform of the ZF-$\mu$SR time spectrum. Solid lines correspond to the local field distributions described by Eq.~\ref{eq:helical_PB} on the four muon sites \#1. See text for details).
}
 \label{fig:HelC}
\end{figure}

The muon time spectrum of MnP at ambient pressure in the helical$-c$ state ($p=0.1$~MPa, $T=20$~K) is presented in Fig.~\ref{fig:HelC}a. The corresponding fast Fourier transform of the $\mu$SR time spectrum is shown in Fig.~\ref{fig:HelC}b. The field distribution presented in Fig.~\ref{fig:HelC}b is characterized by a minimum ($B_{\rm min}$) and a maximum ($B_{\rm max}$) cutoff field, which is consistent with the incommensurate helimagnetic order, and is generally described by the field distribution given by:\cite{Schenk_PRB_2001}
\begin{equation}
P(B)=\frac{2}{\pi}\frac{B}{\sqrt{(B^2-B^2_{\rm min})(B^2_{\rm max}-B^2)}},
 \label{eq:helical_PB}
\end{equation}

Following Refs.~\onlinecite{Felcher_JAP_1966, Obara_JPSJ_1980, Forsyth_PPS_1966, Matsuda_Arxiv_2016}, in the helical$-c$ state the Mn moments are coupled into pairs (Mn1/Mn4 and Mn2/Mn3), and rotate within the $ab-$plane (helical plane) with a constant phase shift between the different pairs [$\alpha=19(5)^{\rm o}$] along the propagation vector ${\bf Q}=(0,0,117)$. The average magnetic moment in the helical state is $m\simeq 1.58$~$\mu_{\rm B}$ with the longer component along the $b-$axis ($m_{b}=1.73$~$\mu_{\rm B}$)  and the shorter one  along the $a-$axis ($m_{a}=1.41$~$\mu_{\rm B}$), respectively.\cite{Forsyth_PPS_1966}

\begin{figure}[htb]
\includegraphics[width=1\linewidth, angle=0]{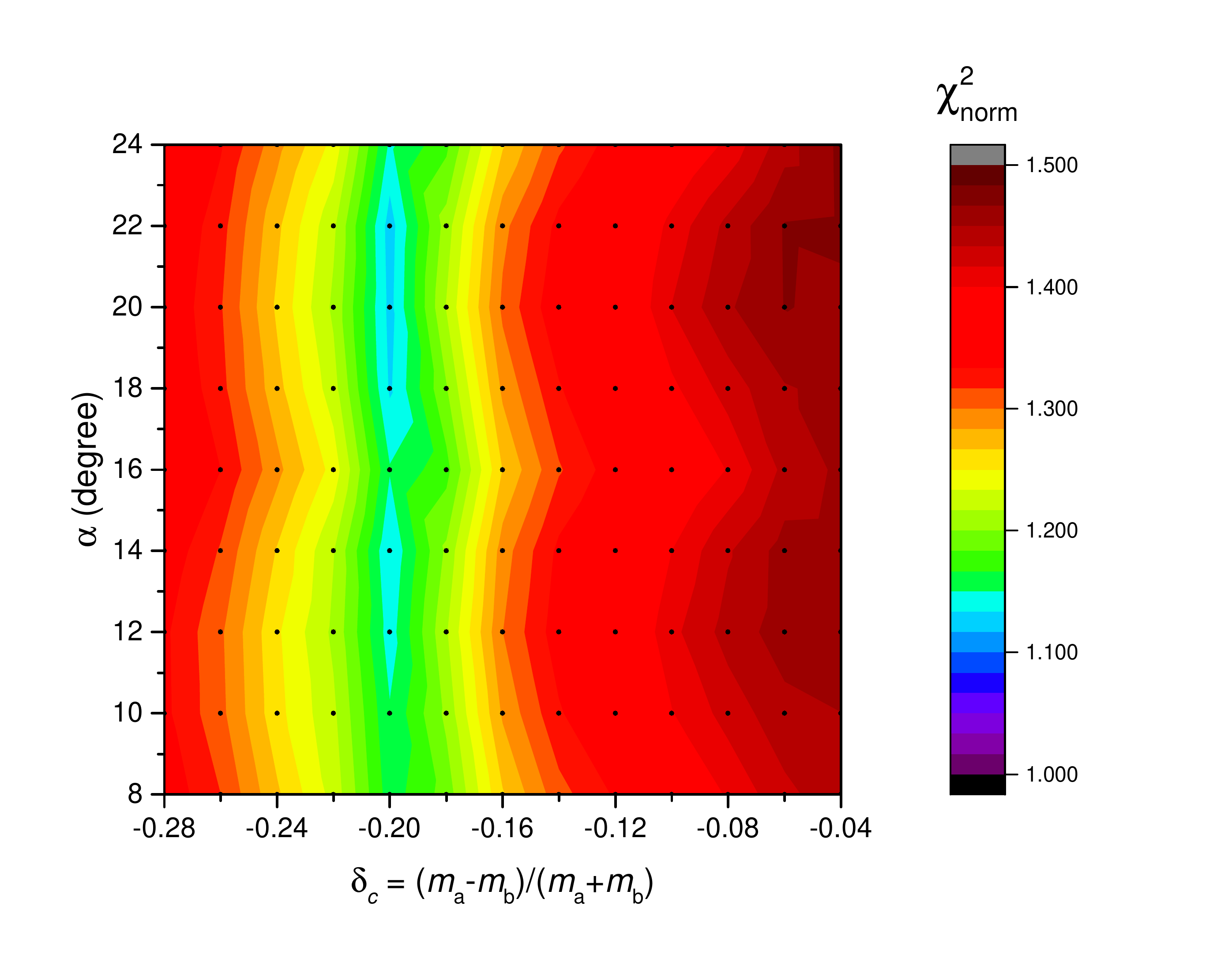}
%
\caption{Correlation plot of $\chi^2_{\rm norm}(\alpha,\delta_c)$ as obtained from the fit of Eq.~\ref{eq:P_Hel} to the experimental ZF-$\mu$SR data presented in Fig.~\ref{fig:HelC}. $\alpha$ is the angle between the two helices (Mn1/Mn4 and Mn2/Mn3) and $\delta_c=(m_a-m_b)/(m_a+m_b)$ is the eccentricity
of the elliptical helical$-c$ structure ($m_a$ and $m_b$ are components of $m$ along the $a-$ and $b-$axes, respectively, see Fig.~\ref{fig:Modulation}). Black dots correspond to set of ($\alpha$, $\delta_c$) points where the calculations and corresponding fits were made.
}
 \label{fig:HelC_chi}
\end{figure}

The comparison of the field distribution given by the helical$-c$ structure with the $\mu$SR data was started by calculating $B_{\rm min}$ and $B_{\rm max}$ for each particular muon site for the phase shift $\alpha$ being in the range of $8^{\rm o}\leq \alpha \leq 24^{\rm o}$ and the eccentricity of the elliptical helical$-c$ structure $\delta_c$ [$\delta_c=(m_a-m_b)/(m_a+m_b)$; $m_a$ and $m_b$ are components of $m$ along the $a-$ and $b-$axes, respectively] being in the range of $-0.28\leq \delta_c \leq -0.04$. The corresponding $\alpha$ and $\delta_c$ values are represented by the black dots in Fig.~\ref{fig:HelC_chi}.
With such determined sets of $B_{\rm min}$ and $B_{\rm max}$ the following function was fitted to the ZF-$\mu$SR time spectra:\cite{Amato_PRB_2014}
\begin{eqnarray}
\frac{A^{\rm Hel}(t)}{A^{\rm Hel}(0)} & = & P^{\rm Hel}(t) =  \frac{1}{3}  \sum_{i=1}^{4}\; w_i e^{-\lambda_{T,i} t} J_0(\gamma_\mu \Delta B_i t)  \label{eq:P_Hel} \\
&&\times\cos(\gamma_\mu B_{av,i}t) +\frac{1}{3}\;e^{-\lambda_L t}. \nonumber
 \end{eqnarray}
Here the index $i$ denotes the $i-$th muon site. $\lambda_i$ is an exponential relaxation rate, $\omega_i$ is the weight ($\omega_i=0.25$ in our case since all four muon sites are equivalent), $\Delta B_i = C\times (B_{\rm max,{\it i}}-B_{\rm min,{\it i}})/2$, and $B_{\rm av,{\it i}}=C\times (B_{\rm max,{\it i}}+B_{\rm min,{\it i}})/2$. The parameter $C\simeq 1$ accounts for the possible deviation of the magnetic moment $m$ from $m=1.58$~$\mu_{\rm B}$ as determined in neutron diffraction experiments by Forsyth {\it et al.}\cite{Forsyth_PPS_1966} Note that according to Eqs.~\ref{eq:Bloc}, \ref{eq:Bdip}, and \ref{eq:Bcont} the local field at the muon stoping site is directly proportional to $m$.

The quality of the fits was checked by using the $\chi^2$ criterium. As shown in Fig.~\ref{fig:HelC_chi} there are two 'local minimum' areas corresponding to $\delta_c\simeq-0.2$  and $\alpha\simeq 13$ or 20$^{\rm o}$ with the normalized $\chi^2_{\rm norm}$ reaching approximately $1.14 - 1.16$. This indicates that the $\mu$SR data are not satisfactorily described by the theory since a good quality fit requires a $\chi^2_{\rm norm}$  to be of the order of unity.

\begin{figure}[htb]
\includegraphics[width=0.6\linewidth, angle=0]{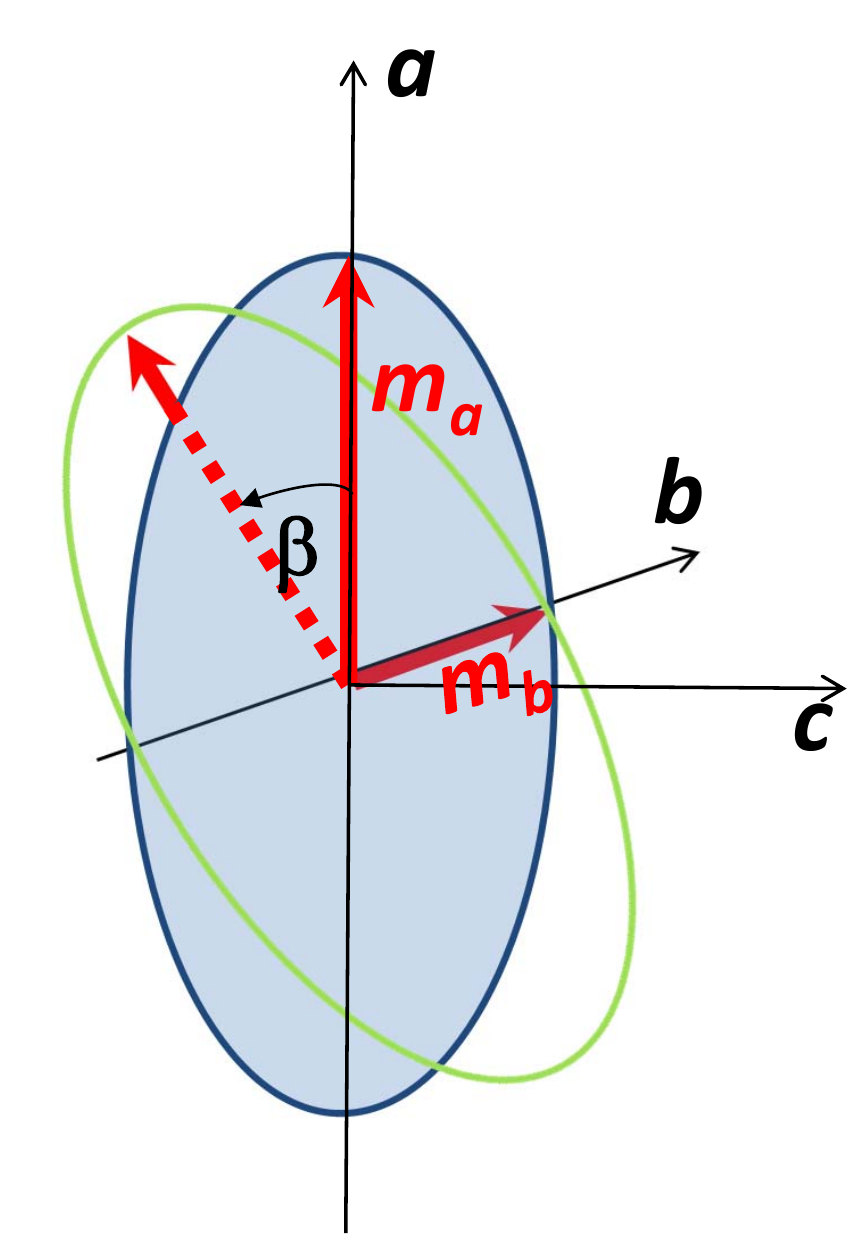}
%
\caption{The tilt of the helical (rotation) plane  on the angle $\beta$  towards the $c-$direction. $m_a$ and $m_b$ are components of the magnetic moment along $a-$ and $b-$axes, respectively.
}
 \label{fig:Modulation}
\end{figure}

Our further study shows that $\chi^2_{\rm norm}$ may be reduced by considering a tilt of the helical (rotation) plane  towards the $c-$direction on the angle $\beta\simeq 4-8^{\rm o}$ (see Fig.~\ref{fig:Modulation}). As suggested in Ref.~\onlinecite{Yamazaki_JPSJ_2014}, the alternatively tilted helimagnetic structure in MnP is indeed stabilized due to Dzyaloshinsky-Moriya interaction and leads to the appearance of an additional modulation of the $a-$component of Mn spins along the $c-$direction (see Fig.~\ref{fig:Modulation}). Considering four Mn spins per unit cell, there are six possibilities for alternative tilts of the rotation planes: $ppmm$, $pmpm$, $pmmp$, $mppm$, $mpmp$, and $mmpp$ [the letter $p$($m$) corresponds to the positive(negative) $\beta$ value; $ppmm$ means {\it e.g.} that $\beta$ is positive for Mn1/Mn2 and negative for Mn3/Mn4 magnetic moments, respectively].

\begin{figure}[htb]
\includegraphics[width=1\linewidth, angle=0]{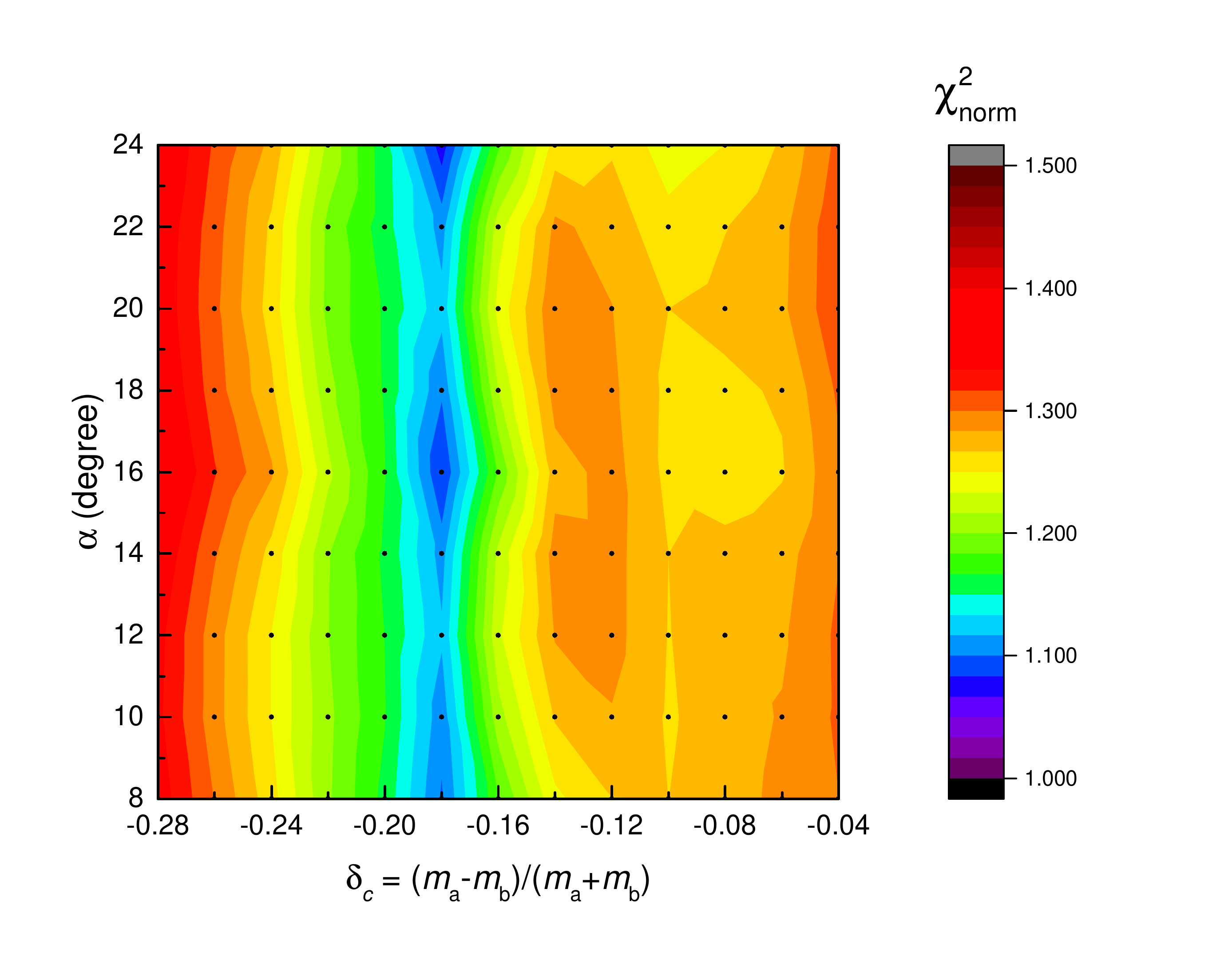}
%
\caption{Correlation plot of $\chi^2_{\rm norm}(\alpha,\delta_c)$ for $\beta=6^{\rm o}$ and $mppm$ helical plane tilt configuration. See text for details. }
 \label{fig:HelC_chi_Mod}
\end{figure}

The analysis reveals that for right-handed helices only the $mppm$ type of tilt leads to a reduction of $\chi^2_{\rm norm}$, while for the rest of tilt configurations the fit results become even worse. For the left-handed helices an improvement of $\chi^2_{\rm norm}$ is observed for the $pmmp$ case. Figure~\ref{fig:HelC_chi_Mod} shows the correlation plot $\chi^2_{\rm norm}(\alpha,\delta_c)$ for right-handed helices in $mppm$ tilt configuration and $\beta = 6^{\rm o}$.
A set of minimum $\chi^2_{\rm norm}$ values is obtained along the $\delta_c\simeq 0.18$ line with approximately $8^{\rm o}$ periodicity ($\alpha\simeq 8^{\rm o}$, $16^{\rm o}$, and $24^{\rm o}$ see Fig.~\ref{fig:HelC_chi_Mod}). The value closest to the one determined by neutron diffraction experiments [$\alpha=19(5)^{\rm o}$] corresponds to the set of parameters: $\alpha\simeq16^{\rm o}$, $\delta_c\simeq 0.18$, and $C=0.982$ and results in $\chi^2_{\rm norm}\simeq 1.08$.
With such defined parameters we get $m_a=1.27$~$\mu_{\rm B}$, $m_b=1.83$~$\mu_{\rm B}$ and $\alpha= 16^{\rm o}$ in reasonable agreement with $m_a=1.41(10)$~$\mu_{\rm B}$, $m_b=1.73(10)$~$\mu_{\rm B}$ and $\alpha=19(5)^{\rm o}$ reported in the literature.\cite{Forsyth_PPS_1966,Felcher_JAP_1966}

It is worth to emphasize, that the averaged Mn moment at each site is estimated to be $\simeq1.55$~$\mu_{\rm B}$, which is larger than 1.29~$\mu_{\rm B}$ determined for the ferromagnetic state. This is similar to the results reported by Forsyth {\it et al.}\cite{Forsyth_PPS_1966}, but contradicts the conclusions of Obara {\it et al.}\cite{Obara_JPSJ_1980} who obtain $m\simeq 1.33$~$\mu_{\rm B}$ in both, the ferromagnetic and the helical$-c$ states, and claim a  'circular' helical$-c$ state with components $m_a=m_b$.

The resulting fit of Eq.~\ref{eq:P_Hel} to the data with $\beta=6^{\rm o}$, $m_a=1.27$~$\mu_{\rm B}$,  $m_b=1.83$~$\mu_{\rm B}$ and $\alpha=16^{\rm o}$ is shown in Fig.~\ref{fig:HelC}a by the solid red line. The corresponding magnetic field distributions for each muon-site obtained by using Eq.~\ref{eq:helical_PB}
are presented in Fig.~\ref{fig:HelC}b.

\section{High-pressure magnetism}\label{sec:high-pressure}

In this Section, we discuss the consistency of the high-pressure magnetic state, which is the precursor of superconductivity, with the helical structures characterized by propagation vectors ${\bf Q} \parallel c$ or ${\bf Q} \parallel b$.
The calculations of the local fields at the muon sites were performed by assuming that the coupling contact constant $A_{\rm cont}\simeq -0.474$~${\rm T}/\mu_{\rm B}$ and the relative positions of Mn atoms and muons  are independent on pressure. The pressure dependencies of the lattice constants $a$, $b$, and $c$ were taken from Ref.~\onlinecite{Wang_Arxiv_2015}.

\begin{figure}[htb]
\includegraphics[width=1\linewidth, angle=0]{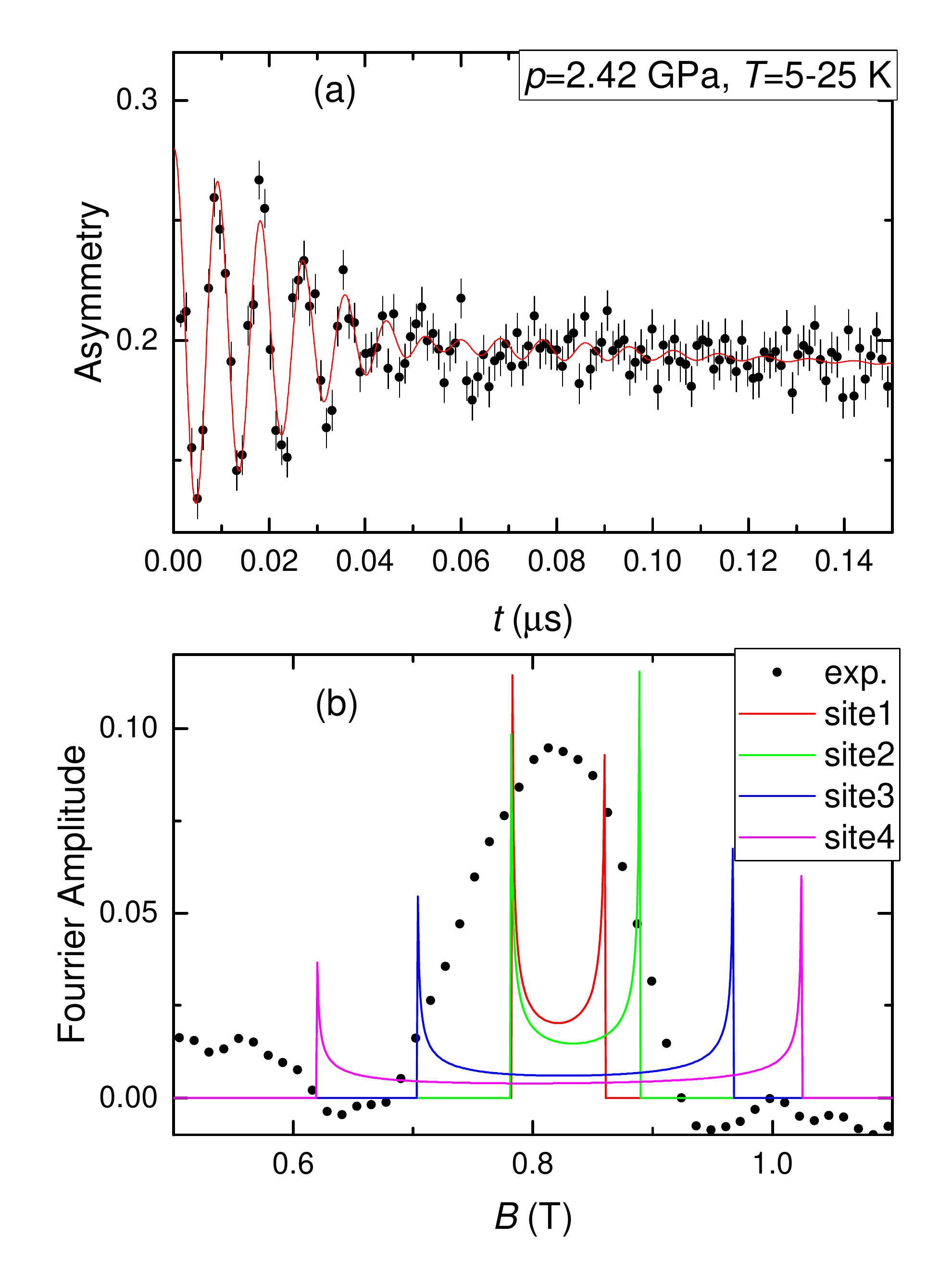}
%
\caption{(a) The ZF-$\mu$SR time spectrum of MnP at $p= 2.42$~GPa.  For increasing accuracy, two time-spectra taken at $T=5$ and 25~K were summed together. The solid line correspond to the fit of Eq.~\ref{eq:P_Hel} to the data for $\alpha=-12^{\rm o}$, $m=1.287$~$\mu_{\rm B}$, and $\delta_b=-0.02$. (b) The fast Fourier transform of the ZF-$\mu$SR time spectrum. Solid lines correspond to the local field distributions described by Eq.~\ref{eq:helical_PB} on four muon sites. See text for details.
}
 \label{fig:HelB}
\end{figure}

The ZF-$\mu$SR time spectrum of MnP at $p=2.42$~GPa is presented in Fig.~\ref{fig:HelB}a. To increase the  accuracy, two time-spectra taken at $T=5$ and 25~K were added. The corresponding fast Fourier transform of the $\mu$SR time spectrum is shown in Fig.~\ref{fig:HelB}b.

\subsection{Consistency of the high-pressure magnetic state with a propagation vector ${\bf Q} \parallel c$ }

Wang {\it et al.}\cite{Wang_Arxiv_2015} have suggested that the pressure induced magnetic order is a helical structure with a  propagation vector ${\bf Q} \parallel c$. The value of ${\bf Q}$ increases from ${\bf Q}=(0,0,0.117)$ at ambient pressure to ${\bf Q}\simeq (0,0,0.25)$ for $p$ exceeding 1.5~GPa. Due to the limited amount of the experimental data the exact magnetic moment arrangement was not identified. The authors of Ref.~\onlinecite{Wang_Arxiv_2015} have only concluded that the constraint between Mn spins leading to coupling of Mn1/Mn4 and Mn2/Mn3 observed at ambient pressure may brake down in the high-pressure state.

In order to check for consistency of the experimental data with a ${\bf Q} \parallel c$ type of helical order, in addition to the angle $\alpha$ (between Mn1/Mn4 and Mn2/Mn3 pairs of spin) the angle $\theta$ [between Mn1 and Mn4 (Mn2 and Mn3) pairs of spins] was introduced. Both $\alpha$ and $\theta$ were allowed to change between $-60^{\rm o}$ and $60^{\rm o}$. The average value of the magnetic moment $m=1.5$~$\mu_{\rm B}$ was taken in accordance with Ref.~\onlinecite{Matsuda_Arxiv_2016}. The magnetic moment eccentricity  $\delta_c=(m_a-m_b)/(m_a+m_b)$ was varied between $-0.5\leq \delta_c \leq 0.5$. The value of the propagation vector ${\bf Q}= (0,0,0.25)$ was kept fixed to the value of Ref.~\onlinecite{Wang_Arxiv_2015}. For each set of the parameters $\alpha$, $\theta$ and $\delta_c$ the corresponding $B_{\rm min,{\it i}}$ $B_{\rm max,{\it i}}$ were obtained by means of Eqs.~\ref{eq:Bloc}, \ref{eq:Bdip} and \ref{eq:Bcont}. With these obtained minimum and maximum fields, Eq.~\ref{eq:P_Hel} was fitted to the experimental data. No agreement between the theory and the experiment was obtained within the above mentioned range of parameters. We conclude, therefore, that the high-pressure helical order with a propagation vector ${\bf Q} \parallel c$ suggested by Wang {\it et al.}\cite{Wang_Arxiv_2015} is inconsistent with our $\mu$SR data.

\subsection{Consistency of the high-pressure magnetic state with a  propagation vector ${\bf Q} \parallel b$}

According to Ref.~\onlinecite{Matsuda_Arxiv_2016}, within the ${\bf Q} \parallel b$ high-pressure helical structure, the Mn moments are coupled in pairs (Mn1/Mn2 and Mn3/Mn4), and rotate within the $ac-$plane with a constant phase difference between the different pairs along ${\bf Q}\sim(0,0.1,0)$. The magnetic moment was found to be elongated along the crystallographic $a-$axis. At $p\simeq 1.8$~GPa and $T=6$~K, ${\bf Q}=(0,0.09,0)$ and the eccentricity  of the moment $\delta_b=(m_a-m_c)/(m_a+m_c)$ is $0.15(8)$.\cite{Matsuda_Arxiv_2016} Hereafter we will call this structure the helical$-b$ structure.

\begin{figure}[htb]
\includegraphics[width=1\linewidth, angle=0]{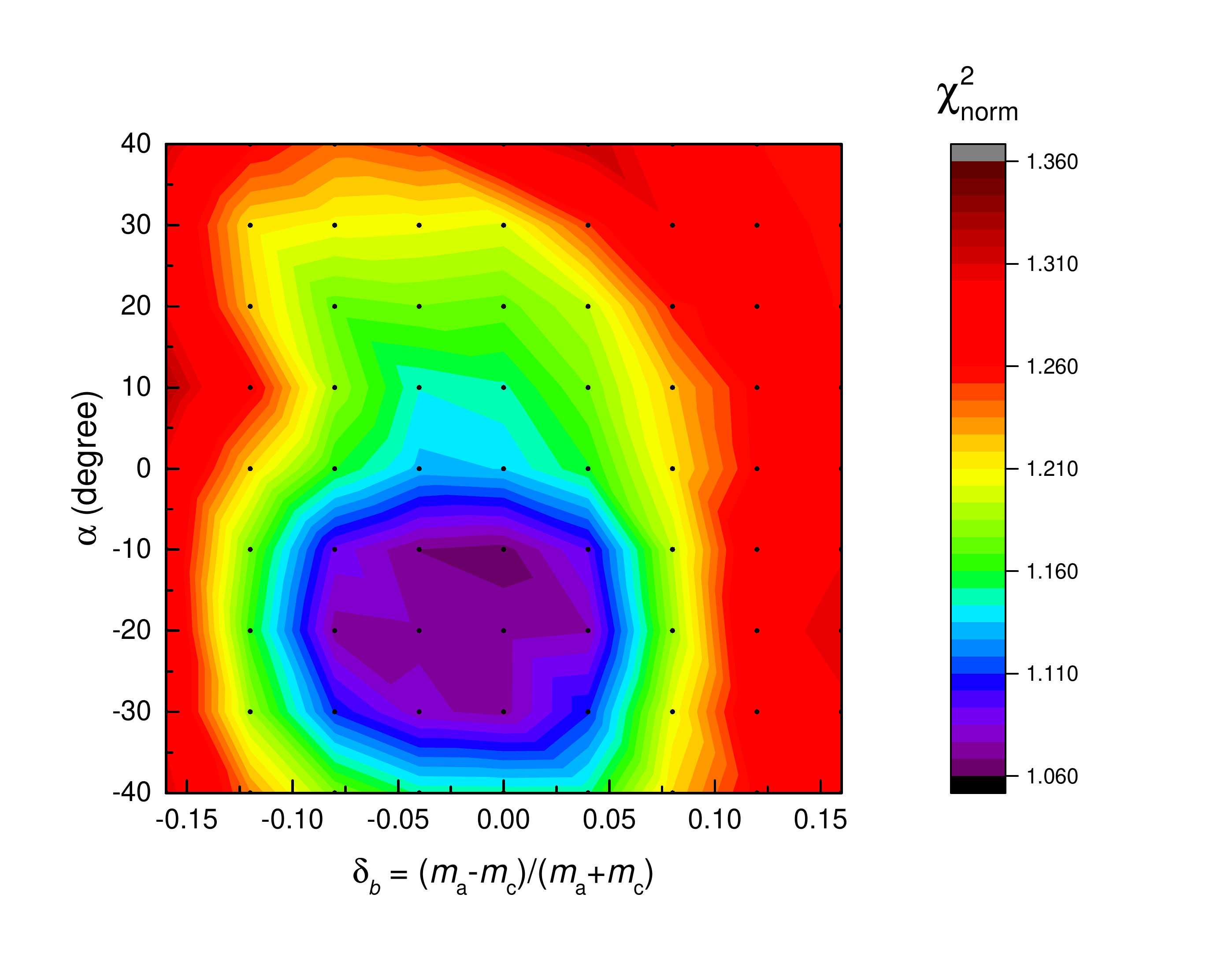}
%
\caption{Correlation plot of $\chi^2_{\rm norm}(\alpha,\delta_b)$ for the double spiral helical$-b$ structure with coupled Mn1/Mn2 and Mn2/Mn3 spin pairs and a propagation vector ${\bf Q}=(0,109,0)$.  See text for details. }
 \label{fig:HelB_chi}
\end{figure}

The extrapolation of Ref.~\onlinecite{Matsuda_Arxiv_2016}'s data gives ${\bf Q}=(0,109,0)$ for $p=2.42$~GPa. The results of our calculations with the above mentioned propagation vector and the parameters $\alpha$ and $\delta_b$ varied from $-40^{\rm o} \leq \alpha \leq 40^{\rm o}$ and $-0.16 \leq \delta_b \leq 0.16$ are presented in Fig.~\ref{fig:HelB_chi}.
A well pronounced minimum with $\chi^2_{\rm norm}\simeq 1.07$ is observed for $\alpha\simeq -12^{\rm o}$ and $\delta_b\simeq -0.02$. The corresponding average magnetic moment value is $m=1.287$~$\mu_{\rm B}$.

The fit of Eq.~\ref{eq:P_Hel} to the data with $\alpha=-12^{\rm o}$, $m=1.287$~$\mu_{\rm B}$, and $\delta_b=-0.02$ is shown in Fig.~\ref{fig:HelB}a by the solid red line. The corresponding magnetic field distributions for each muon-site obtained by using Eq.~\ref{eq:helical_PB} are presented in Fig.~\ref{fig:HelB}b.

\subsection{Consistency of the high-pressure magnetic state with a  propagation vector ${\bf Q} \parallel b$ and $c-$axis modulation}

In contrast to the results of Wang {\it at al.}\cite{Wang_Arxiv_2015} no diffraction peaks at $(0,0,1\pm0.25)$ were observed by Matsuda {\it et al.}\cite{Matsuda_Arxiv_2016} in neutron diffraction experiments. This allows them to conclude that the ''signals $(0,0,1\pm0.25)$ may be sample dependent or come from the surface``.\cite{Matsuda_Arxiv_2016} The fact that our high-pressure $\mu$SR data are inconsistent with ${\bf Q} \parallel c$ helical order may rule out the sample dependent issue.

\begin{figure}[htb]
\includegraphics[width=1\linewidth, angle=0]{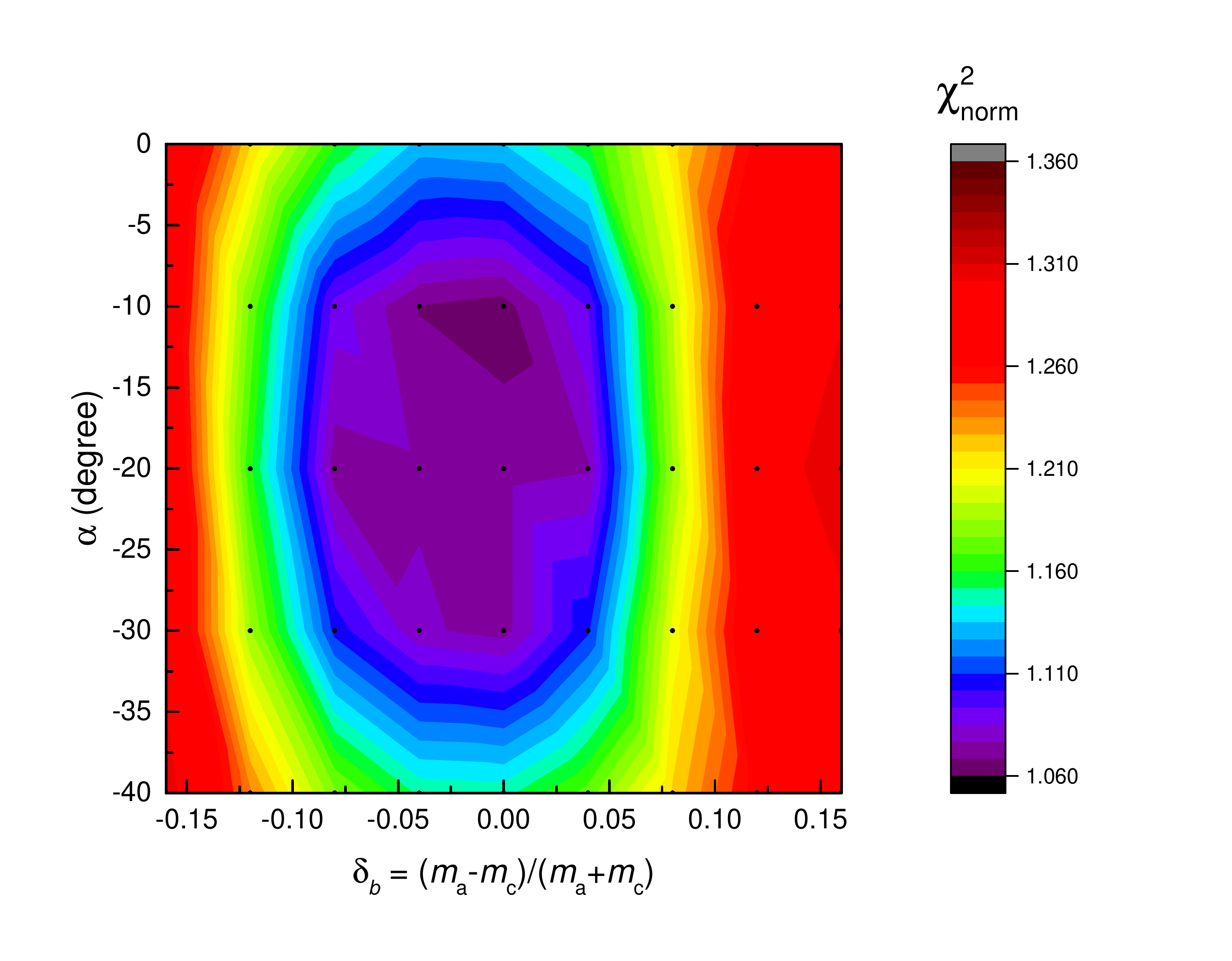}
%
\caption{Correlation plot of $\chi^2_{\rm norm}(\alpha,\delta_b)$ for 10$^{rm o}$ $mppm$ tilt configuration. See text for details.
}
 \label{fig:HelB_chi_Mod}
\end{figure}

There is, probably, another explanation allowing to link together the results presented in Refs.~\onlinecite{Matsuda_Arxiv_2016} and \onlinecite{Wang_Arxiv_2015}. In analogy with the ambient pressure data  (see Ref.~\onlinecite{Yamazaki_JPSJ_2014} and Sec.~\ref{sec:HelC}) we may assume a possible tilt of the $ac$ rotation plane towards $b-$direction leading to the appearance of a modulation of the $c-$component of  Mn spins along the crystallographic $b-$axis.
By following the discussion in Sec.~\ref{sec:HelC} the 'tilted` helimagnetic structures with six possible alternative tilt of the rotation planes, namely $ppmm$, $pmpm$, $pmmp$, $mppm$, $mpmp$, and $mmpp$, were considered. The analysis reveal no substantial change in correlation $\chi^2_{\rm norm}(\alpha,\delta_b)$ plots for {\it all} six configurations with the tilt angle growing up to at least $10^{\rm o}$. As an example, Fig.~\ref{fig:HelB_chi_Mod} shows the $\chi^2_{\rm norm}(\alpha,\delta_b)$ plot for the $mppm$ configuration and a $10^{\rm o}$ tilting angle. Still, the well pronounced minima with $\chi^{2}_{norm}\simeq 1.07$ is observed for $\alpha\simeq-12^{\rm o}$, and $\delta_b\simeq-0.02$, thus suggesting that the modulated helical$-b$ structure remains consistent with our $\mu$SR high-pressure data.

It should be mentioned, however, that within this simple approach the period of modulation of the magnetic moment along $b-$axis is uniquely determined by the propagation vector [$Q=(0,0.109,0)$ in our case]. This leads to appearance of additional $(0,0,1\pm 0.109)$ peaks but not $(0,0,1\pm0.25)$ peaks as reported in Ref.~\onlinecite{Wang_Arxiv_2015}. It could be, therefore, that either more complicated modulations of helical planes take place in MnP at high pressures or, as suggested in Ref.~\onlinecite{Matsuda_Arxiv_2016}, the observation of Wang {\it et al.}\cite{Wang_Arxiv_2015} correspond to the surface only.

\section{Conclusions}\label{sec:conclusions}

The muon-spin rotation data collected at ambient pressure and $p=2.42$~GPa in MnP were checked for consistency  with various magnetic structures reported up to date in the literature.\cite{Felcher_JAP_1966, Obara_JPSJ_1980, Forsyth_PPS_1966, Yamazaki_JPSJ_2014, Wang_Arxiv_2015, Matsuda_Arxiv_2016, Khasanov_PRB_2016}

At ambient pressure two phases with very different $\mu$SR responses are clearly detected. In the ferromagnetic state the Mn spins are all aligned along the crystallographic $b-$direction. The comparison of our $\mu$SR data with the known FM Mn spin arrangement allows to obtain the coupling contact constant $A_{\rm cont}=-0.474$~T/$\mu_{\rm B}$. At low-temperatures      a double spiral helimagnetic structure is observed. The elliptical helical structure with ${\bf Q}=(0,0,0.117)$, the $a-$axis moment elongated by approximately 18\% and the rotation plane tilted towards $c-$direction on the angle $\simeq 4-8^{\rm o}$ leads to a reasonably good agreement between the theory and the experiment. The phase difference between Mn1/Mn4 and Mn2/Mn3 sets of spins was estimated to be $\simeq 16^{\rm o}$. The determined averaged Mn moment is 1.55~$\mu_{\rm B}$, which is higher than the 1.29~$\mu_{\rm B}$ in the FM state.

The analysis of the high-pressure $\mu$SR data reveal that the new magnetic order appearing for pressures exceeding $1.5$~GPa can not be described by keeping the direction of the propagation vector unchanged. Even the extreme case -- decoupling the double-helical structure into four individual helices -- remains inconsistent with the experiment. The experimental data are well described with the double spiral helimagnetic structure with the propagation vector ${\bf Q}=(0,0.109,0)$ and the $c-$axis moment elongated by approximately 2\% in comparison with the $a-$axis one. The phase difference between Mn1/Mn2 and Mn3/Mn4 sets of moments and the value of the average magnetic moment $m$ were estimated to be $\simeq -12^{\rm o}$ and 1.287~$\mu_{\rm B}$, respectively.
The agreement between the contradicting neutron and $\mu$SR data on the one side,\cite{Matsuda_Arxiv_2016,Khasanov_HPR_2016} and the x-ray data on the other side,\cite{Wang_Arxiv_2015} might be reached by assuming the tilt of the rotation plane towards $b-$direction.

The work was performed at the Swiss Muon Source (S$\mu$S), PSI, Villigen. The work of ZG was supported  by the Swiss National Science Foundation (SNF-Grant 200021-149486). PB thanks the computing resources provided by CINECA within the Scientific Computational Project CINECA ISCRA Class C (Award HP10C5EHG5, 2015) and STFC's Scientific Computing Department. The work PB and RdR was supported by the grants from MUON JRA of EU FP7 NMI3, under grant agreement 226507.

\end{document}